\pdfoutput=1
\documentclass[aps, pra, showpacs, showkeys,
twocolumn
]{revtex4}

\usepackage{amsmath}
\usepackage{graphicx}
\usepackage{color}

\usepackage{hyperref}

\usepackage{microtype}      % enables character protrusion and font expansion

\hypersetup{
colorlinks=true,
linkcolor=blue,
citecolor=blue,
urlcolor=blue,
bookmarks,
breaklinks,
pdftitle={Generating entangled fermions by accelerated measurements on the
vacuum},
pdfauthor={David C. M. Ostapchuk, Robert B. Mann},
pdfsubject={PACS numbers: 03.67.Bg, 04.62.+v, 05.30.Fk},
pdfkeywords={entanglement generation, Davies-Unruh effect, quantum information,
quantum field theory, curved spacetime}
}

\newcommand{\ket}[1]{\ensuremath{\lvert#1\rangle}}      % ket

\newcommand{\rrw}{\ensuremath{I}}           % right Rindler wedge
\newcommand{\lrw}{\ensuremath{{I\!I}}}      % left Rindler wedge

\DeclareMathOperator{\Tr}{Tr}           % trace
\DeclareMathOperator{\diag}{diag}       % diagonal matrix

\newcommand{\mpc}[1]{\ensuremath{a_{#1}^\dagger}}   % creation
\newcommand{\mpa}[1]{\ensuremath{a_{#1}}}           % annihilation
\newcommand{\mpket}[1]{\ensuremath{\ket{#1}^+}}     % Fock state

\newcommand{\mac}[1]{\ensuremath{b_{#1}^\dagger}}   % creation
\newcommand{\maa}[1]{\ensuremath{b_{#1}}}           % annihilation 
\newcommand{\maket}[1]{\ensuremath{\ket{#1}^-}}     % Fock state

\newcommand{\ripc}[1]{\ensuremath{c^{\rrw\dagger}_{#1}}}    % creation
\newcommand{\ripa}[1]{\ensuremath{c^\rrw_{#1}}}             % annihilation 
\newcommand{\ripket}[1]{\ensuremath{\ket{#1}_\rrw^+}}       % Fock state

\newcommand{\riac}[1]{\ensuremath{d^{\rrw\dagger}_{#1}}}    % creation
\newcommand{\riaa}[1]{\ensuremath{d^\rrw_{#1}}}             % annihilation
\newcommand{\riaket}[1]{\ensuremath{\ket{#1}_\rrw^-}}       % Fock state

\newcommand{\riipc}[1]{\ensuremath{c^{\lrw\dagger}_{#1}}}   % creation
\newcommand{\riipa}[1]{\ensuremath{c^{\lrw}_{#1}}}          % annihilation
\newcommand{\riipket}[1]{\ensuremath{\ket{#1}_\lrw^+}}      % Fock state

\newcommand{\riiac}[1]{\ensuremath{d^{\lrw\dagger}_{#1}}}   % creation
\newcommand{\riiaa}[1]{\ensuremath{d^{\lrw}_{#1}}}          % annihilation
\newcommand{\riiaket}[1]{\ensuremath{\ket{#1}_\lrw^-}}      % Fock state

\begin{document}

\title{Generating entangled fermions by accelerated measurements on the vacuum}

\author{David C. M. Ostapchuk}
\email{dcmostap@uwaterloo.ca}

\author{Robert B. Mann}
\email{rbmann@uwaterloo.ca}

\affiliation{Department of Physics and Astronomy, University of Waterloo,
Waterloo, Canada, N2L 3G1}
\affiliation{Institute for Quantum Computing, University of Waterloo, Waterloo,
Canada, N2L 3G1}

\date{October 8, 2008}


\begin{abstract}
\phantomsection
\addcontentsline{toc}{section}{\protect\numberline{}\abstractname}
It is shown that accelerated projective measurements on the vacuum of a free
Dirac spinor field results in an entangled state for an inertial observer. The
physical mechanism at work is the Davies-Unruh effect. The produced state is
always entangled and its entanglement increases as a function of the
acceleration, reaching maximal entanglement in the asymptotic limit of infinite
acceleration where Bell states are produced.
\end{abstract}

\maketitle

Recently, much attention has been given to relativistic effects in the context
of quantum information theory. Since relativity is an indispensable component of
any complete theoretical model, understanding these effects is important from
the viewpoint of fundamental physics. However, it could also be relevant in
practical situations in which quantum information processing tasks are
implemented by observers in arbitrary relative motion.

Entanglement plays a pivotal role in quantum information --- it is a resource
for quantum communication and teleportation and for various computational
tasks~\cite{Nielsen2000Quantum-Computa}. In a relativistic setting, the role
played by entanglement has received much recent
attention~\cite{Peres2004Quantum-informa, Alsing2002Lorentz-Invaria,
Gingrich2002Quantum-Entangl, Pachos2003Generation-and-, Kim2005Lorentz-invaria,
Alsing2003Teleportation-w, Alsing2004Teleportation-i,
Fuentes-Schuller2005Alice-falls-int, Adesso2007Continuous-vari,
Alsing2006Entanglement-of, Ball2006Entanglement-in}.  While it has been shown to
be an invariant quantity for observers in uniform relative
motion~\cite{Peres2004Quantum-informa, Alsing2002Lorentz-Invaria,
Gingrich2002Quantum-Entangl, Pachos2003Generation-and-, Kim2005Lorentz-invaria},
in non-inertial frames the situation is quite different.

The fidelity of teleportation between two parties in relative uniform
acceleration decreases as a function of
acceleration~\cite{Alsing2003Teleportation-w, Alsing2004Teleportation-i,
Schutzhold2005Comment-on-Tele}.  From the perspective of a uniformly
accelerating observer a communication horizon appears, obstructing access to
information about the whole of spacetime.  As a consequence, there is a loss of
information and a corresponding degradation of entanglement. This has been shown
to hold for both scalar fields~\cite{Fuentes-Schuller2005Alice-falls-int,
Adesso2007Continuous-vari} and spinor fields~\cite{Alsing2006Entanglement-of} as
viewed by two relatively accelerating observers. The acceleration of the
observer effectively introduces an ``environmental decoherence'' that limits the
fidelity of certain quantum information-theoretic processes. Implementation of
quantum information processing tasks between accelerating partners thus depends
upon a proper and quantitative understanding of such degradation in non-inertial
frames.

In this paper we analyze the generation of entanglement between different modes
of a Dirac spinor field due to projective measurements on the vacuum by an
accelerating observer. For any observer the vacuum is the absence of both
particles and antiparticles as measured by that observer's detector. As a result
of the Davies-Unruh
effect~\cite{Davies1975Scalar-particle,Unruh1976Notes-on-black-}, an
accelerating observer will perceive a Fermi-Dirac distribution of particles and
antiparticles in what an inertial observer would describe as the vacuum state.
We show that if one of these particles is detected, an entangled state is
produced in the inertial reference frame. We show that entanglement is always
produced and quantify it using the entanglement entropy. We find that larger
accelerations produce more entanglement and in the asymptotic limit of infinite
acceleration, a maximally entangled Bell state is produced. A similar effect
holds for scalar fields~\cite{Han2008Generating-Enta}; however, further
processing is required to extract a Bell state, even in the asymptotic limit of
infinite acceleration. We will work in units where $c = \hbar = k_B = 1$.

The Davies-Unruh effect for a Dirac spinor field $\Psi$ of mass $m$ is a
consequence of two inequivalent quantization
schemes~\cite{Fulling1973Nonuniqueness-o, Birrell1982Quantum-Fields-}. For the
inertial observer in flat spacetime, the appropriate metric is the Minkowski
metric $g_{\mu\nu} = \eta_{\mu\nu} = \diag(1,-1,-1,-1)$. Since this metric is
static, the field can be quantized in a straightforward manner by expanding it
in terms of a complete set of positive and negative frequency modes (suppressing
henceforth the spin degree of freedom for ease of notation)
% Eqn: Field expansion in Minkowski modes
\begin{equation*}
\Psi = \int dk \, ( \mpa{k} \, \psi^+_k + \mac{k} \, \psi^-_k ),
\end{equation*}
and imposing the canonical anticommutation relations on the mode operators
$\{ \mpa{k},\mpc{k^\prime} \}=\{ \maa{k},\mac{k^\prime} \}=\delta(k-k^\prime)$,
with all other anticommutators vanishing.

The key element here is the division of the modes into positive and negative
frequency, which is done according to the Minkowski timelike Killing vector
$\partial_t$. The operators $\mpc{k}$ and $\maa{k}$ are then interpreted as
particle creation operators and antiparticle annihilation operators,
respectively. With this interpretation, a Fock space can be constructed in the
usual manner.

\begin{figure}
\centering\includegraphics{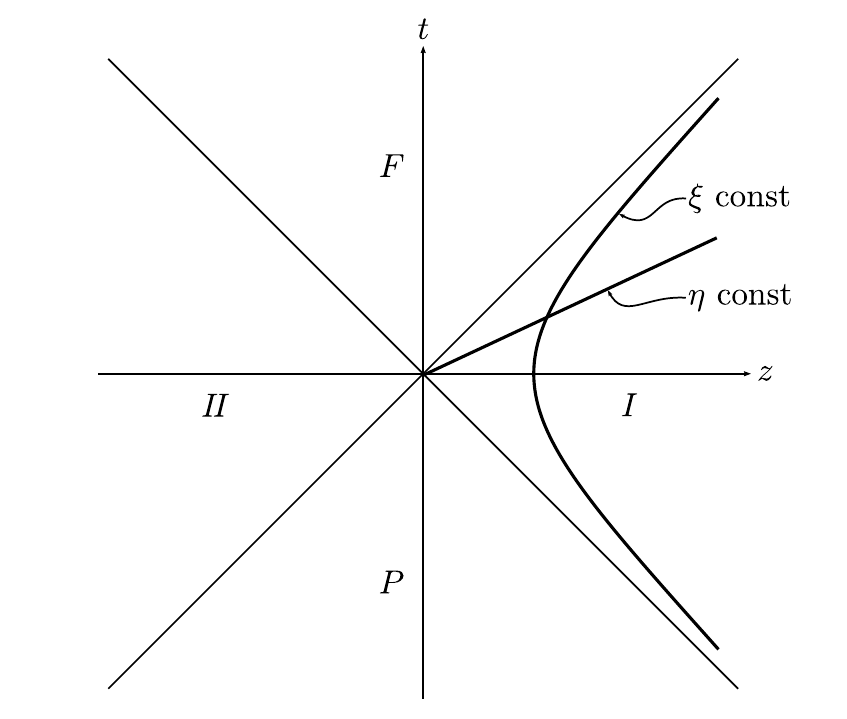}
\caption{Spacetime is naturally divided into four regions denoted $\rrw$, $F$,
$\lrw$, $P$ for a uniformly accelerating observer. The line $\xi = 0$ is the
worldline of the observer and $\eta$ is his or her proper time.}
\label{fig:rindler}
\end{figure}

Now consider an observer moving through flat spacetime with uniform acceleration
$a$ in the $z$ direction. This observer will experience communication horizons
that divide the spacetime into four regions denoted $\rrw$, $\lrw$, $F$, and $P$
(see Fig.~\ref{fig:rindler}). The observer will be confined to region $\rrw$,
which is causally disconnected from region $\lrw$. The appropriate coordinates
to describe his or her motion are the Rindler
coordinates~\cite{Rindler1966Kruskal-Space-a}, given by
% Eqn: Rindler coordinates (right wedge)
\begin{align*}
at & = e^{a\xi} \, \sinh{a\eta}, & az & = e^{a\xi} \, \cosh{a\eta},
\end{align*}
with the remaining coordinates unchanged. The coordinates take the values
$-\infty < \eta, \xi < \infty$ and cover region $\rrw$. A separate Rindler
coordinate system is needed to cover region $\lrw$ and it differs from the above
by an overall minus sign. Both coordinates give rise to the same line element
$ds^2 = \exp(2a\xi) \, ( d\eta^2 - d\xi^2 )$. In these coordinates, the line
$\xi=0$ is the worldline of the accelerating observer and $\eta$ is his or her
proper time.

% Fig: Rindler spacetime

The quantum field theory for the Rindler observer is constructed by expanding
the field in terms of the complete set of positive and negative frequency
Rindler modes
% Eqn: Field expansion in Rindler modes
\begin{equation*}
\Psi = \int dk \, ( \ripa{k} \, \psi^{\rrw+}_k + \riac{k} \, \psi^{\rrw-}_k +
\riipa{k} \, \psi^{\lrw+}_k + \riiac{k} \, \psi^{\lrw-}_k ),
\end{equation*}
and imposing the canonical anticommutation relations on the mode operators
% Eqn: Rindler canonical anticommutation relations
\begin{align*}
\{ \ripa{k}, \ripc{k^\prime} \} & = \{ \riaa{k}, \riac{k^\prime}\}
= \delta(k - k^\prime), \\
\{ \riipa{k}, \riipc{k^\prime} \} & = \{ \riiaa{k}, \riiac{k^\prime} \}
= \delta(k - k^\prime),
\end{align*}
with all other anticommutators vanishing.

There are two types of Rindler modes to reflect the causal structure of Rindler
spacetime: the modes $\psi^{\rrw\pm}_k$ have support in region $\rrw$ whereas
the modes $\psi^{\lrw\pm}_k$ have support in region $\lrw$. Each type is divided
into positive and negative frequency according to the Rindler timelike Killing
vector in the appropriate region. In region $\rrw$ this is given by
$\partial_\eta$, however, in region $\lrw$ it is $\partial_{-\eta}$ where the
minus sign ensures it is future pointing.

The operators $\ripa{k}$ and $\riac{k}$ annihilate a particle and create an
antiparticle in region $\rrw$ while $\riipa{k}$ and $\riiac{k}$ annihilate and
create particles and antiparticles in region $\lrw$.

These two quantization schemes are not
equivalent~\cite{Fulling1973Nonuniqueness-o}.  Making the single mode
approximation, in which the Rindler observer's particle detector is sensitive to
a narrow bandwidth centered about the perpendicular components of the wavevector
$\vec{k}_\perp$ (which is the same for a Minkowski
observer)~\cite{Alsing2004Teleportation-i, Alsing2003Teleportation-w,
Alsing2006Entanglement-of}, the mode operators are related by the following
Bogoliubov transformation
% Eqn: Bogoliubov transformation: Minkowski particle annihilation operator
\begin{equation}
\label{eq:bogompa}
\begin{pmatrix}
    \mpa{k} \\
    \mac{-k}
\end{pmatrix}
=
\begin{pmatrix}
\cos{r_k} & -e^{-i\phi_k} \sin{r_k} \\
e^{i\phi_k} \sin{r_k} & \cos{r_k}
\end{pmatrix}
\begin{pmatrix}
    \ripa{k} \\
    \riiac{-k}
\end{pmatrix},
\end{equation}
where the parameter $r_k$ is defined by
% Eqn: Definition of $r_k$
\begin{equation}
\label{eq:rk}
\cos{r_k} = [2\cosh(\pi\omega_k/a)]^{-1/2} \, \exp(\pi \omega_k/2a),
\end{equation}
and $\omega_k = \sqrt{k^2 + \vec{k}_\perp^2 + m^2}$ is the frequency of the
mode. The phase $\phi_k$ can be absorbed into the definitions of the mode
operators and will be done so from now on using the sign conventions
of~\cite{Alsing2006Entanglement-of}.

Using these transformations, the Minkowski particle vacuum in mode $k$ can be
expressed in terms of Rindler Fock states as
% Eqn: Minkowski particle vacuum in Rindler
\begin{equation}
\label{eq:mpvac}
\mpket{0_k} = \cos{r_k} \, \exp \bigl( \tan(r_k) \, \ripc{k} \, \riiac{-k}
\bigr) \ripket{0_k} \riiaket{0_{-k}},
\end{equation}
where the $+$ ($-$) superscripts denote particle (antiparticle). A formal
expression for the total Minkowski particle vacuum is obtained by using
Eq.~\eqref{eq:mpvac} for each mode
% Eqn: Total Minkowski particle vacuum in Rindler
\begin{equation}
\label{eq:minkpvac}
\mpket{0} = N \prod_{k} \exp \bigl( \tan(r_k) \, \ripc{k} \, \riiac{-k} \bigr)
\ripket{0_k} \riiaket{0_{-k}},
\end{equation}
where $N = \prod_k \cos{r_k}$.

Similarly, if the Rindler observer's antiparticle detector is sensitive to a
narrow bandwidth centered about the wavevector $-\vec{k}_\perp$ then
% Eqn: Bogoliubov transformation: Minkowski antiparticle annihilation operator
\begin{equation}
\label{eq:bogomaa}
\begin{pmatrix}
    \maa{k} \\
    \mpc{-k}
\end{pmatrix}
=
\begin{pmatrix}
\cos{r_k} & e^{-i\phi_k} \sin{r_k} \\
-e^{i\phi_k} \sin{r_k} & \cos{r_k}
\end{pmatrix}
\begin{pmatrix}
    \riaa{k} \\
    \riipc{-k}
\end{pmatrix},
\end{equation}
and the Minkowski antiparticle vacuum for mode $k$ takes the form
% Eqn: Minkowski antiparticle vacuum in Rindler
\begin{equation*}
\maket{0_k} = \cos{r_k} \, \exp \bigl( -\tan(r_k) \, \riac{k} \, \riipc{-k}
\bigr) \riaket{0_k} \riipket{0_{-k}}.
\end{equation*}
Formally, the total Minkowski antiparticle vacuum is then
% Eqn: Total Minkowski antiparticle vacuum in Rindler
\begin{equation*}
\maket{0} = N \prod_{k} \exp \bigl( -\tan(r_k) \, \riac{k} \, \riipc{-k}, \bigr)
\riaket{0_k} \riipket{0_{-k}},
\end{equation*}
with the full Minkowski vacuum being $\ket{0} = \mpket{0}\maket{0}$,
corresponding to the absence of particles and antiparticles as detected by the
Minkowski observer.

We now see that the accelerating observer has a nonzero probability to detect
particles and antiparticles in the Minkowski vacuum. The probabilities are given
according to the Fermi-Dirac distribution of temperature $T = a/2\pi$, which can
be seen by constructing the reduced density matrix for region~$\rrw$.

What are the consequences of detecting one of these particles? Suppose two
observers, Alice, an inertial observer, and Rob, a uniformly accelerating
observer, are moving through the field $\psi$. When the field is in the vacuum
state as described by Alice, Rob would describe the state as the thermal
state~\eqref{eq:minkpvac} due to the Davies-Unruh effect. Now suppose Rob
performs a measurement on this state and detects one particle in mode $k$.
Immediately after his measurement, the state will be the projection
of~\eqref{eq:minkpvac} onto the single particle state in region~$\rrw$. This can
be written succinctly as
% Eqn: State after the detection of a particle
\begin{equation*}
\ket{\psi_+(k)} = P_{k} N \prod_{k^\prime} \exp \bigl( \tan(r_{k^\prime}) \,
\ripc{k^\prime} \, \riiac{-k^\prime} \bigr) \ripket{0_{k^\prime}}
\riiaket{0_{-k^\prime}},
\end{equation*}
where the operator $P_k$ is defined as
% Eqn: Definition of $P_k$
\begin{equation*}
P_{k} = \sec(r_k) \ripc{k} \, \riiac{-k} \exp \bigl( -\tan(r_k) \, \ripc{k} \,
\riiac{-k} \bigr).
\end{equation*}
Applying the Bogoliubov transformation~\eqref{eq:bogompa}, the state can be
simplified to
% Eqn: State after particle detection in Minkowski
\begin{equation}
\label{eq:minkpsik}
\ket{\psi_+(k)} = \bigl[ \sin(r_k) + \cos(r_k) \, \mpc{k} \, \mac{-k} \bigr]
\ket{0},
\end{equation}
from which we see that from Alice's perspective, the state is a superposition of
the vacuum (i.e., no particle emission) and pair production at energy
$\omega_k$. This state is entangled in the occupation number of the particle
mode $k$ and the antiparticle mode $-k$.

\begin{figure}
\centering\includegraphics{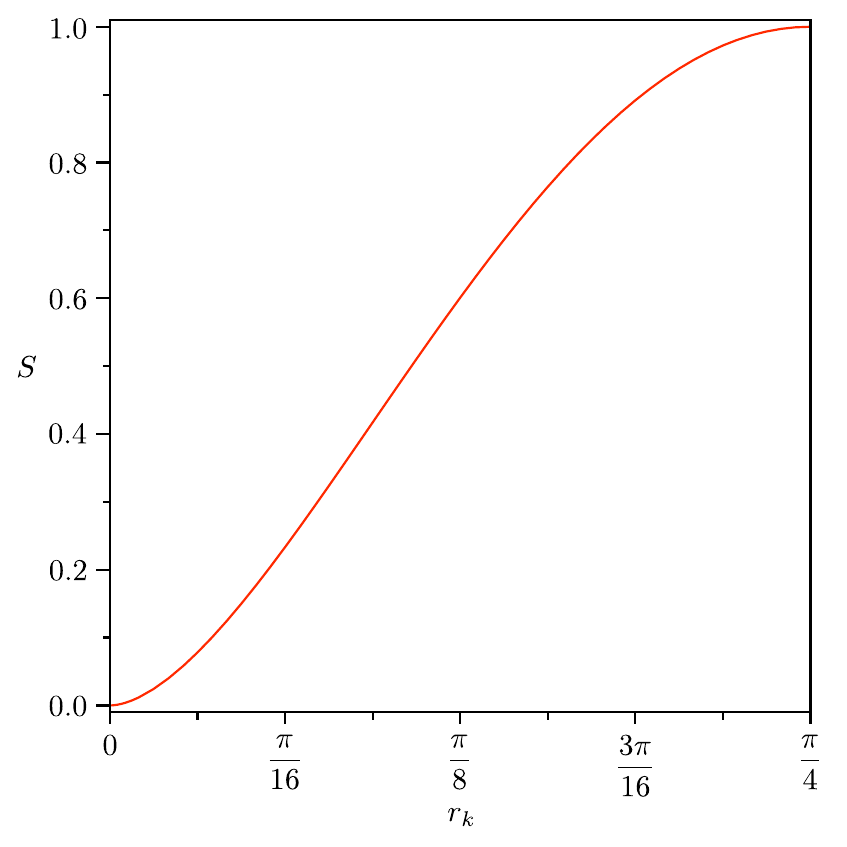}
\caption{Entanglement entropy of the produced state~\eqref{eq:minkpsik} as a
function of $r_k$. Larger accelerations produce more entanglement reaching the
maximum of $1$ when $r_k = \pi/4$.}
\label{fig:entropy}
\end{figure}

To study the entanglement properties of this state we work in the basis
$\{\ket{\tilde{0}}^+, \ket{\tilde{1}}^+, \lvert\tilde{0}\rangle^-,
\ket{\tilde{1}}^-\}$ where $\ket{\tilde{0}}^\pm = \ket{0_{\pm k}}^\pm$ and
$\ket{\tilde{1}}^\pm = \ket{1_{\pm k}}^\pm$. The state can then be represented
by the density matrix
% Eqn: Density matrix of state after particle detection
\begin{equation*}
\rho(k) =
\begin{pmatrix}
\sin^2{r_k} & 0 & 0 & \sin{r_k} \, \cos{r_k} \\
0 & 0 & 0 & 0 \\
0 & 0 & 0 & 0 \\
\sin{r_k} \, \cos{r_k} & 0 & 0 & \cos^2{r_k}
\end{pmatrix}.
\end{equation*}
We quantify the entanglement of the state in terms of the entanglement entropy,
defined as the von Neumann entropy~\cite{Nielsen2000Quantum-Computa} $S(\rho_a)
= -\Tr(\rho_a \log_2{\rho_a})$ of $\rho_a$, the reduced density matrix of
subsystem $a$. Equivalently, it can be expressed in terms of the eigenvalues
$\lambda_i$ of $\rho_a$ as
% Eqn: Definition of von Neumann entropy in terms of eigenvalues
\begin{equation*}
S(\rho_a) = -\sum_i \lambda_i \log_2{\lambda_i}.
\end{equation*}
For a pure bipartite state, it does not matter which subsystem is traced out as
the nonzero eigenvalues of either reduced density matrix are equal. To find the
entanglement entropy of~\eqref{eq:minkpsik}, we find the reduced density matrix
by tracing out the particle states to obtain
% Eqn: Reduced density matrix of state after particle detection
\begin{equation*}
\rho_-(k) = \Tr_+ \rho(k) =
\begin{pmatrix}
\sin^2{r_k} & 0 \\
0 & \cos^2{r_k}
\end{pmatrix}.
\end{equation*}
From which we calculate the entropy to be
% Eqn: Entanglement entropy of state after particle detection
\begin{equation*}
S\bigl(\rho_-(k) \bigr) = \log_2( \csc^2{r_k} ) + \cos^2(r_k) \log_2(
\tan^2{r_k} ).
\end{equation*}
Recalling that $r_k$ is defined by~\eqref{eq:rk}, we see that the entanglement
entropy is nonzero regardless of the frequency detected or the (nonzero)
acceleration of the observer. Therefore, the state always contains distillable
entanglement with larger accelerations producing more entanglement, as
illustrated in Fig.~\ref{fig:entropy}.

% Fig: Plot of entanglement entropy

Note that in order for Alice to use this state in a quantum information
processing task, she must know Rob's acceleration and the momentum of the
particle detected. Rob can communicate these parameters to Alice classically
since he can always signal to her despite there being a point where he can no
longer receive signals from her.  It is interesting to note that if Alice does
not know Rob's acceleration, she may be able to deduce it from the resulting
quantum state. We leave this for future work.

In the limit $\omega_k/a \rightarrow 0$ we have $r_k = \pi/4$ and the state
approaches the maximally entangled Bell state
% Eqn: Asymptotic limit of infinite acceleration
\begin{equation*}
\lim_{\omega_k/a \rightarrow 0} \ket{\psi_+(k)} = \frac{1}{\sqrt{2}}(
\mpket{\tilde{0}} \maket{\tilde{0}} + \mpket{\tilde{1}} \maket{\tilde{1}} ),
\end{equation*}
which has an entanglement entropy of $1$. This limit corresponds physically to
the asymptotic limit of infinite acceleration. However, whenever
$\omega_k \ll a$, the state is approximately
% Eqn: Approximate state for small $\omega_k/a$
\begin{widetext}
\begin{equation*}
\ket{\psi_+(k)} \approx \sqrt{2} \left( \frac{1}{2} - \frac{\pi\omega_k}{4a} -
\frac{\pi^2\omega_k^2}{16a^2} \right) \ket{\tilde{0}}^+ \ket{\tilde{0}}^- +
\sqrt{2} \left( \frac{1}{2} - \frac{\pi\omega_k}{4a} +
\frac{\pi^2\omega_k^2}{16a^2}\right) \ket{\tilde{1}}^+ \ket{\tilde{1}}^-,
\end{equation*}
\end{widetext}
which has an entanglement entropy of 
% Eqn: Approximate entanglement entropy for small $\omega_k/a$
\begin{equation*}
S( a, \omega_k) = 1 - \frac{\pi^2 \omega_k^2}{2\ln(2)a^2} + O\left(
\frac{\omega_k^4}{a^4}  \right).
\end{equation*}
In the case of massless fermions, entanglement arbitrarily close to maximal can
be generated for finite acceleration by detecting sufficiently low energy modes.
However, this is not true in the massive case where accelerations at least much
greater than the rest mass energy are required to approximate a Bell state.

While the above analysis is conditioned on Rob detecting a single particle in
mode $k$, it generalizes to other measurement outcomes. If he had instead
detected an antiparticle in mode $k$, the resulting state would be
% Eqn: State after the detection of an antiparticle
\begin{equation*}
\ket{\psi_-(k)} = A_k N \prod_{k^\prime} \exp \bigl( -\tan(r_{k^\prime}) \,
\riac{k^\prime} \, \riipc{-k^\prime} \bigr) \riaket{0_{k^\prime}}
\riipket{0_{-k^\prime}},
\end{equation*}
where the operator $A_k$ is defined as
% Eqn: Definition of $A_k$
\begin{equation*}
A_k = -\sec(r_k) \riac{r_k} \, \riipc{r_-k} \exp \bigl( \tan(r_k) \riac{r_k} \,
\riipc{r_{-k}} \bigr).
\end{equation*}
Upon applying the Bogoliubov transformation~\eqref{eq:bogomaa} this simplifies
to
% Eqn: State after antiparticle detection in Minkowski
\begin{equation*}
\ket{\psi_-(k)} = \bigl[ \sin(r_k) - \cos(r_k) \mac{k} \, \mpc{-k} \bigr]
\ket{0},
\end{equation*}
which also approaches a Bell state in the asymptotic limit of infinite
acceleration.

Noting that the operators $P_k$ and $A_{k^\prime}$ only contain an even number
of mode operators, they will commute. Therefore, the state after an arbitrary
measurement result will be the product of the states $\ket{\psi_\pm(k)}$ for
each mode detected. Physically, this would be a superposition of all possible
pair productions including no pair production; in the asymptotic limit of
infinite acceleration, this approaches a product of Bell states. Regardless of
the acceleration, given this state, Alice could use it as a resource in quantum
information processing tasks.

In summary, we have shown that accelerated projective measurements on the
Minkowksi vacuum of a free Dirac spinor field produce entangled states for
inertial observers.  These could in principle be used in quantum information
processing tasks. The produced states are always entangled and the degree to
which they are entangled is a function of the frequency of the detected
particles and the acceleration of the observer. The amount of entanglement
increases with acceleration and reaches maximal entanglement in the asymptotic
limit of infinite acceleration where a product of Bell states is produced.

%%%%%%%%%%%%%%%%%%%%%%%%%%%%%%%%%%%%%%%%%%%%%%%%%%%%%%%%%%%%%%%%%%%%%%%%%%%%%%%%
% Acknowledgements
%%%%%%%%%%%%%%%%%%%%%%%%%%%%%%%%%%%%%%%%%%%%%%%%%%%%%%%%%%%%%%%%%%%%%%%%%%%%%%%%

\begin{acknowledgments}
We would like to thank Ivette Fuentes-Schuller, Jorma Louko, and Iman Marvian
for useful discussion and offering valuable feedback.  This work was supported
in part by the Natural Sciences and Engineering Research Council of Canada.
\end{acknowledgments}

\end{document}